\documentclass[aps,prb,onecolumn,twoside,floatfix]{revtex4}
\usepackage{amsmath}
\usepackage{mathtools}
\usepackage{amssymb}
\usepackage{graphicx}
\usepackage{dcolumn}
\usepackage{bm}

\usepackage[]{breqn}
\renewcommand\[{\begin{dmath}}
\renewcommand\]{\end{dmath}}
\usepackage{float}
\makeatletter
\let\cat@comma@active\@empty
\makeatother

\begin{document}


\title[Short-range order in high-entropy alloys: \\
Theoretical formulation and application to Mo-Nb-Ta-V-W system ]{Short-range order in high entropy alloys: \\
Theoretical formulation and application to Mo-Nb-Ta-V-W system }

\author{
A. Fern\'andez-Caballero,
}

\affiliation{
CCFE, United Kingdom Atomic Energy Authority, Abingdon , Abingdon, OX14 3DB, UK
}
\affiliation{ 
School of Mechanical Aerospace and Civil Engineering, University of Manchester, UK
}
\author{
J.S. Wr\'obel,
}
\affiliation{
Faculty of Materials Science and Engineering, Warsaw University of Technology, Woloska 141 02-507 Warsaw Poland
}
\author{
P.M.  Mummery, 
}
\affiliation{
School of Mechanical Aerospace and Civil Engineering, University of Manchester, UK
}
\author{
	D. Nguyen-Manh,
}
\email{
Duc.Nguyen@ukaea.uk
}
\affiliation{
CCFE, United Kingdom Atomic Energy Authority, Abingdon , Abingdon, OX14 3DB, UK
}

\date{\today}

\begin{abstract}
In high-entropy alloys (HEAs), the local chemical fluctuations from disordered solute solution state into segregation, precipitation and ordering configurations are complex due to the large number of elements.  In this work, the cluster expansion (CE) Hamiltonian for multi-component alloy systems is developed in order to investigate the dependence of chemical ordering of HEAs as a function of temperature dependence due to derivation of configuration entropy from the ideal solute solution. Analytic expressions for Warren-Cowley short-range order (SRO) parameters are derived for a five component alloy system. The theoretical formulation is used to investigate the evolution of the ten different SRO parameters in the MoNbTaVW and the sub-quaternary systems obtained by Monte-Carlo simulations within the combined CE and first-principles formalism.  The strongest chemical SRO parameter is predicted for the first nearest-neighbor Mo-Ta pair that is in consistent agreement with high value of enthalpy of mixing in the $B2$ structure for this binary system. The prediction of $B2$ phase presence for Mo-Ta pairs in the considered bcc HEAs is reinforced by the positive contribution to the average SRO from the second nearest-neighbor shell. Interestingly, it is found that the average SRO parameter for the first and second nearest-neighbor shells of V-W pairs is also strongly negative in a comparison with the Mo-Ta pairs. This finding in the HEAs can be rationalized and discussed by the presence of the ordered-like  $B32$ phase which has been predicted as the ground-state structure in binary bcc V-W system at the equimolar composition.     
\end{abstract}

\maketitle


\section{\label{sec:level1}Introduction}
Multi-principal element alloys, termed High Entropy Alloys (HEAs) with predominantly single solid solution phases have become an emerging field fo\emph{}r alloy development with many basic concepts, including the origin of entropy effect on interplay between thermodynamic analysis of complex, concentrated alloys and their microstructural properties [\onlinecite{Miracle2017,Pickering2016a,Zhang2014}]. This new class of materials, first brought to the attention in 2004 through the work of Cantor et al. [\onlinecite{Cantor2004}] and Yeh et al. [\onlinecite{Yeh2004}], is based around the concept that their high configurational entropy of mixing should stabilize simple solid-solution phases (such as fcc or bcc) relative to the formation of potentially-embrittling intermetallic ones. Recent experimental investigations cast doubts about the effect of entropic stabilization of solid solutions in different HEAs [\onlinecite{Gali2013,Pickering2016b,Leong2017,Calvo2017}]. In particular, the formation of two distinct types of Cr-rich precipitates ($M_{23}C_{6}$ and the $\sigma$ phases) has been reported in the initially believed single fcc phase of the HEAs CrMnFeCoNi following long-term heat treatment [\onlinecite{Pickering2016b}]. 

Nevertheless, HEAs offer not only new and exciting approach to alloy design but also attract interest due to the discovery of alloys with unusual and attractive properties including physical, chemical, magnetic and mechanical properties [\onlinecite{Zhang2014,Calvo2017}]. An example of exceptional mechanical property is having greater yield strength than any of its individual constituents in the bcc-structured refractory high entropy alloys MoNbTaVW. These HEAs have high melting points and the excellent yield strength sustained to ultrahigh temperatures has been usually attributed to solid solute strengthening mechanism and/or associated lattice strains [\onlinecite{Senkov2010}]. In general, the HEAs are characterized not only by high values of entropy but also by high atomic-level stresses originating from mixing of elements with different sizes [\onlinecite{Toda-Caraballo2015,Kozak2015}].  Therefore, there are still many fundamental questions that need to be addressed to understand local atomic structure in the terms order and disorder of HEAs. Each element in HEAs will tend to occupy the position that minimizes its site energy and bond energies which in turn depend on preferred local chemical environments, inter-atomic interactions and atomic volumes of different constituent species. To experimentally determine the elemental distribution, several complementary techniques have been applied to ascertain the HEAs local structure include neutron scattering, high energy synchroton x-ray diffraction, atom probe tomography and transmission electron microscopy coupled with XEDS (X-ray energy dispersive spectroscopy)[\onlinecite{Calvo2017,Santodonato2015,Williams2014}].

In recent years, HEAs have attracted significant attention due to their superior radiation resistance compared to conventional single phase Fe-Cr-Ni austenitic stainless steels making them potential candidates for high-temperature fission and fusion applications [\onlinecite{Kumar2016,Granberg2016}]. Stability upon irradiation after cascade events may be attributed to presence of high atomic level stresses resulted from difference in atomic sizes in HEAs and their tendency for amorphization and recrystallization at high rates after irradiation induced thermal spike [\onlinecite{Egami2014}]. It is worth mentioning that W based alloys are considered the preferred option for plasma facing materials in magnetically confined fusion reactor designs[\onlinecite{Stork2014}]. Beside the high melting point, the reasons for the interest are high thermal conductivity, low activation, low tritium retention and low sputtering yield with regards to radiation damage for structural materials in nuclear fusion power plants [\onlinecite{Zinkle2009,NguyenManh2012}]. 

In the present work, a predictive model developed previously for investigating phase stability in multi-component alloys [\onlinecite{Toda-Caraballo2015,Wrobel2015}] has been  employed in order to quantify short-range ordering of different atomic species for HEAs from first-principles calculations in a combination with the Cluster Expansion (CE) approach. The  CE Hamiltonian can be used to describe both enthalpy and entropy contributions to free energy consistently not only for solid solution but also inter-metallic phases in multi-component alloy systems. In particular, by using thermodynamic integration via Monte-Carlo simulations with the effective cluster interactions (ECIs), it is shown that in general the configurational entropy contribution depends strongly on temperature and therefore the entropy expression for ideal random solid solution phenomenologically adopted for HEAs is valid only at high-temperature limit[\onlinecite{Wrobel2015}]. The non-random configuration gives a tendency toward phase separation or chemical short-range ordering (SRO) and both of these trends decrease the configurational entropy from ideal estimates. This paper is organized as follows. In section II, the CE formalism is introduced and developed for the Hamiltonian of multi-component system with a special focus on quinary alloys. In section III, the mathematical expression of short-range order formulas are derived explicitly for five element alloy in terms of average pair correlation functions. In section IV, the SRO expressions are applied to the short range order parameters for the specific HEAs MoNbTaVW and the corresponding five quaternary sub-systems. The second nearest-neighbor contributions into SRO properties are discussed in section V. A summary of our results is given in section VI.


\section{Cluster expansion Hamiltonian for multicomponent alloys}
The phase stability of multi-component alloys can be investigated using a combination of Density Functional Theory (DFT) technique and lattice statistical simulations based on CE formalism [\onlinecite{Toda-Caraballo2015,Wrobel2015}]. DFT calculations were performed using the projector augmented wave (PAW) method implemented in Vienna Ab-initio Simulation Package (VASP) [\onlinecite{PhysRevB.47.558,PhysRevB.49.14251,KRESSE199615,PhysRevB.54.11169,PhysRevB.50.17953,PhysRevB.59.1758}]. We use PAW potentials with semi-core p electron contribution with 11 electrons treated as valence for V, Nb, Ta and 12 electrons for Mo and W. Exchange and correlation interactions were treated in the generalized gradient approximation GGA-PBE [\onlinecite{Perdew1996}]. 

The enthalpy of mixing for a five-component alloy, which can be evaluated using DFT, can be defined as 

\[
\text{$\Delta $H}_{\text{mix}} ( \vec{\sigma } )=\text{E}_{\text{tot}}^{\text{lat}} \left(A_{c_A}B_{c_B}C_{c_C}D_{c_D}E_{c_E},\vec{\sigma }\right)-c_A \text{E}_{\text{tot}}^{\text{lat}} \text{(A)}-c_B \text{E}_{\text{tot}}^{\text{lat}} \text{(B)}-c_C \text{E}_{\text{tot}}^{\text{lat}} \text{(C)}-c_D \text{E}_{\text{tot}}^{\text{lat}} \text{(D)}-c_E \text{E}_{\text{tot}}^{\text{lat}} \text{(E)}
\label{eq:basic_enthalpy_mix}
\]

where $c_{A}$, $c_{B}$, $c_{C}$, $c_{D}$, $c_{E}$ are the average concentrations of alloy components A, B, C, D and E, respectively, and $E_{total}^{lattice}$ is the total energy per atom for considered structure. Here the vector $\vec{\sigma }$ defines the alloy configuration for a given lattice such as body-centered cubic (bcc). Within the cluster expansion formalism [\onlinecite{vandeWalle2002,0965-0393-10-5-304,vandeWalle2009266,Hart2008}], the configurational  enthalpy of mixing from Eq.(\ref{eq:basic_enthalpy_mix}) can be expressed in term of different cluster interaction energies by the following formula 

\[
\text{$\Delta $H}_{\text{mix}} ( \vec{\sigma } )=\sum _{\omega }^{\text{}} J_{\omega } m_{\omega }  \left\langle \Gamma_{\omega'}(\vec{\sigma})\right\rangle_\omega
\label{eq:CE}
\]
 
where the summation in Eq.(\ref{eq:CE}) is performed over all the clusters $\omega$ distinct under symmetry operations within the underlying lattice. $m_{\omega}$ denotes the multiplicities indicating the number of clusters equivalent to $\omega$ by symmetry and $J_{\omega }$ are the effective cluster interactions corresponding to cluster $\omega$. $\left\langle \Gamma_{\omega'}(\vec{\sigma})\right\rangle_\omega$ are the cluster functions defined as product of point functions of occupation variables, $\gamma_{i}(\sigma_{p})$, on specific cluster $\omega$ and averaged over the clusters of atoms $\omega'$ that are equivalent to cluster $\omega$. The general expression for the cluster correlation function corresponding to an alloy configuration given by $\vec{\sigma}$ is given by:

\[
\left\langle \Gamma _{|\omega|,m}^{\text{(ijk$\cdots $)}} ( \vec{\sigma } ) \right\rangle =\sum _{\text{pqr$\cdots $}} \gamma _{\text{i}} ( \sigma _{\text{p}} ) \gamma _{\text{j}} ( \sigma _{\text{q}} ) \gamma _{\text{k}} ( \sigma _{\text{r}} ) \cdots  \text{y}_m^{\text{pqr$\cdots $}}
\label{eq:general_corr_function}
\]

Here m is an integer corresponding to the configuration of the atomic labellings of the lattice points in the cluster shell. $\text{y}_m^{\text{pqr$\cdots$ }}$ denotes the temperature-dependent probability of finding atomic species $\text{pqr$\cdots$ }$ which are located at atomic configuration in the shell specified by m. For a general n-component systems, the orthogonal point functions $\gamma$, are defined as in [\onlinecite{vandeWalle2009266}] by:
\begin{equation}
\gamma_{j,n}\left(\sigma_i\right)=\begin{cases}
1 & \textrm{ if }j=0\textrm{ }, \\
-\cos\left(2\pi\lceil\frac{j}{2}\rceil\frac{\sigma_i}{n}\right) & \textrm{ if }j>0\textrm{ and odd},  \\
-\sin\left(2\pi\lceil\frac{j}{2}\rceil\frac{\sigma_i}{n}\right) & \textrm{ if }j>0\textrm{ and even},
\end{cases}
\label{eq:gamma}
\end{equation}

From Eqs.(\ref{eq:general_corr_function}) and (\ref{eq:gamma}), the cluster correlation functions for point and pair clusters are given, respectively, by:

\begin{equation}
\left\langle \Gamma _{1,m}^{\text{(i)}} ( \vec{\sigma } ) \right\rangle =\sum _p \gamma _{\text{i}} ( \sigma _{\text{p}} ) \text{y}_m^p=\sum _p \gamma _{\text{i}} ( \sigma _{\text{p}} ) \text{c}_p
\label{eq:point_corr_function}
\end{equation}
\[
\left\langle \Gamma _{2,m}^{\text{(ij)}} ( \vec{\sigma } ) \right\rangle =\sum _{\text{pq}} \gamma _{\text{i}} ( \sigma _{\text{p}} ) \gamma _{\text{j}} ( \sigma _{\text{q}} ) \text{y}_{m}^{\text{pq}}
\label{eq:pair_corr_function}
\]

Note that in Eq.(\ref{eq:point_corr_function}), the average single-site cluster functions can be expressed in term of alloy concentration $c_p$. For the average pairwise cluster functions, Eq.(\ref{eq:pair_corr_function}),  the index m denotes the $m^{th}$ nearest-neighbor pair which has probability $\text{y}_m^{\text{pq}}$ with atom of type p sitting at the first site and atom of type q at the second site of the $m^{th}$ pair.

The expression for the single-site correlation function for five-component alloy can be obtained by using Eq.(\ref{eq:point_corr_function}) in terms of average atomic concentration of all the elements present in the system as follows

\begin{widetext}
\begin{subequations}
\[
\left\langle \Gamma _{1,1}^0 \right\rangle=1
\]
\[
\left\langle \Gamma _{1,1}^1 \right\rangle=\frac{1}{4} \left[\phi _- \left(\text{c}_{\text{B}}+\text{c}_{\text{E}}\right)+\phi _+ \left(\text{c}_{\text{C}}+\text{c}_{\text{D}}\right)-4 \text{c}_{\text{A}}\right]
\]
\[
\left\langle \Gamma _{1,1}^2 \right\rangle=\sqrt{\chi _-} \left(\text{c}_{\text{D}}-\text{c}_{\text{C}}\right)+\sqrt{\chi _+} \left(\text{c}_{\text{E}}-\text{c}_{\text{B}}\right)
\]
\[
\left\langle \Gamma _{1,1}^3 \right\rangle=\frac{1}{4} \left[\phi _- \left(\text{c}_{\text{C}}+\text{c}_{\text{D}}\right)+\phi _+ \left(\text{c}_{\text{B}}+\text{c}_{\text{E}}\right)-4 \text{c}_{\text{A}}\right]
\]
\[
\left\langle \Gamma _{1,1}^4 \right\rangle=\sqrt{\chi _-} \left(\text{c}_{\text{E}}-\text{c}_{\text{B}}\right)+\sqrt{\chi _+} \left(\text{c}_{\text{C}}-\text{c}_{\text{D}}\right)
\]
\label{eq:5_component_point_corr_functions}
\end{subequations}

where we used the following notation: $\phi _\pm=1\pm\sqrt{5}$, $\chi _\pm=\frac{5}{8}\pm\frac{\sqrt{5}}{8}$. From Eq.(\ref{eq:pair_corr_function}), the expressions for the pairwise cluster functions can be expressed in terms of the pair probabilities, $\text{y}_m^{\text{pq}}$, as follows:
\begin{subequations}
\[
\left\langle \Gamma _{\text{2,m}}^{11} \right\rangle=\frac{1}{16} \left[2 \phi _- \phi _+ \left(\text{y}_{\text{m}}^{\text{BC}}+\text{y}_{\text{m}}^{\text{BD}}+\text{y}_{\text{m}}^{\text{CE}}+\text{y}_{\text{m}}^{\text{DE}}\right)-8 \phi _- \left(\text{y}_{\text{m}}^{\text{AB}}+\text{y}_{\text{m}}^{\text{AE}}\right)+\phi _-^2 \left(\text{y}_{\text{m}}^{\text{BB}}+2 \text{y}_{\text{m}}^{\text{BE}}+\text{y}_{\text{m}}^{\text{EE}}\right)-8 \phi _+ \left(\text{y}_{\text{m}}^{\text{AC}}+\text{y}_{\text{m}}^{\text{AD}}\right)+\phi _+^2 \left(\text{y}_{\text{m}}^{\text{CC}}+2 \text{y}_{\text{m}}^{\text{CD}}+\text{y}_{\text{m}}^{\text{DD}}\right)+16 \text{y}_{\text{m}}^{\text{AA}}\right]
\]
\[
\left\langle \Gamma _{\text{2,m}}^{12} \right\rangle=\frac{1}{4} \left[\phi _- \sqrt{\chi _+} \left(\text{y}_{\text{m}}^{\text{EE}}-\text{y}_{\text{m}}^{\text{BB}}\right)+\sqrt{\chi _-} \phi _+ \left(\text{y}_{\text{m}}^{\text{DD}}-\text{y}_{\text{m}}^{\text{CC}}\right)+4 \sqrt{\chi _-} \left(\text{y}_{\text{m}}^{\text{AC}}-\text{y}_{\text{m}}^{\text{AD}}\right)+\sqrt{\chi _-} \phi _- \left(-\text{y}_{\text{m}}^{\text{BC}}+\text{y}_{\text{m}}^{\text{BD}}-\text{y}_{\text{m}}^{\text{CE}}+\text{y}_{\text{m}}^{\text{DE}}\right)+4 \sqrt{\chi _+} \left(\text{y}_{\text{m}}^{\text{AB}}-\text{y}_{\text{m}}^{\text{AE}}\right)+\sqrt{\chi _+} \phi _+ \left(-\text{y}_{\text{m}}^{\text{BC}}-\text{y}_{\text{m}}^{\text{BD}}+\text{y}_{\text{m}}^{\text{CE}}+\text{y}_{\text{m}}^{\text{DE}}\right)\right]
\]
\[
\left\langle \Gamma _{\text{2,m}}^{13} \right\rangle=\frac{1}{16} \left[\phi _- \phi _+ \left(\text{y}_{\text{m}}^{\text{BB}}+2 \text{y}_{\text{m}}^{\text{BE}}+\text{y}_{\text{m}}^{\text{CC}}+2 \text{y}_{\text{m}}^{\text{CD}}+\text{y}_{\text{m}}^{\text{DD}}+\text{y}_{\text{m}}^{\text{EE}}\right)-4 \phi _- \left(\text{y}_{\text{m}}^{\text{AB}}+\text{y}_{\text{m}}^{\text{AC}}+\text{y}_{\text{m}}^{\text{AD}}+\text{y}_{\text{m}}^{\text{AE}}\right)+\phi _-^2 \left(\text{y}_{\text{m}}^{\text{BC}}+\text{y}_{\text{m}}^{\text{BD}}+\text{y}_{\text{m}}^{\text{CE}}+\text{y}_{\text{m}}^{\text{DE}}\right)-4 \phi _+ \left(\text{y}_{\text{m}}^{\text{AB}}+\text{y}_{\text{m}}^{\text{AC}}+\text{y}_{\text{m}}^{\text{AD}}+\text{y}_{\text{m}}^{\text{AE}}\right)+\phi _+^2 \left(\text{y}_{\text{m}}^{\text{BC}}+\text{y}_{\text{m}}^{\text{BD}}+\text{y}_{\text{m}}^{\text{CE}}+\text{y}_{\text{m}}^{\text{DE}}\right)+16 \text{y}_{\text{m}}^{\text{AA}}\right]
\]
\[
\left\langle \Gamma _{\text{2,m}}^{14} \right\rangle=\frac{1}{4} \left[\phi _- \sqrt{\chi _+} \left(\text{y}_{\text{m}}^{\text{BC}}-\text{y}_{\text{m}}^{\text{DE}}\right)+\sqrt{\chi _-} \phi _+ \left(\text{y}_{\text{m}}^{\text{DE}}-\text{y}_{\text{m}}^{\text{BC}}\right)+\left(\phi _- \sqrt{\chi _+}+\sqrt{\chi _-} \phi _+\right) \left(\text{y}_{\text{m}}^{\text{CE}}-\text{y}_{\text{m}}^{\text{BD}}\right)+4 \sqrt{\chi _-} \left(\text{y}_{\text{m}}^{\text{AB}}-\text{y}_{\text{m}}^{\text{AE}}\right)+\sqrt{\chi _-} \phi _- \left(\text{y}_{\text{m}}^{\text{EE}}-\text{y}_{\text{m}}^{\text{BB}}\right)-4 \sqrt{\chi _+} \left(\text{y}_{\text{m}}^{\text{AC}}-\text{y}_{\text{m}}^{\text{AD}}\right)+\sqrt{\chi _+} \phi _+ \left(\text{y}_{\text{m}}^{\text{CC}}-\text{y}_{\text{m}}^{\text{DD}}\right)\right]
\]
\[
\left\langle \Gamma _{\text{2,m}}^{22} \right\rangle=2 \sqrt{\chi _- \chi _+} \left(\text{y}_{\text{m}}^{\text{BC}}-\text{y}_{\text{m}}^{\text{BD}}-\text{y}_{\text{m}}^{\text{CE}}+\text{y}_{\text{m}}^{\text{DE}}\right)+\chi _- \left(\text{y}_{\text{m}}^{\text{CC}}-2 \text{y}_{\text{m}}^{\text{CD}}+\text{y}_{\text{m}}^{\text{DD}}\right)+\chi _+ \left(\text{y}_{\text{m}}^{\text{BB}}-2 \text{y}_{\text{m}}^{\text{BE}}+\text{y}_{\text{m}}^{\text{EE}}\right)
\]
\[
\left\langle \Gamma _{\text{2,m}}^{23} \right\rangle=\frac{1}{4} \left[\left(\phi _- \sqrt{\chi _+}+\sqrt{\chi _-} \phi _+\right) \left(\text{y}_{\text{m}}^{\text{DE}}-\text{y}_{\text{m}}^{\text{BC}}\right)+\phi _- \sqrt{\chi _+} \left(\text{y}_{\text{m}}^{\text{CE}}-\text{y}_{\text{m}}^{\text{BD}}\right)+\sqrt{\chi _-} \phi _+ \left(\text{y}_{\text{m}}^{\text{BD}}-\text{y}_{\text{m}}^{\text{CE}}\right)+4 \sqrt{\chi _-} \left(\text{y}_{\text{m}}^{\text{AC}}-\text{y}_{\text{m}}^{\text{AD}}\right)+\sqrt{\chi _-} \phi _- \left(\text{y}_{\text{m}}^{\text{DD}}-\text{y}_{\text{m}}^{\text{CC}}\right)+4 \sqrt{\chi _+} \left(\text{y}_{\text{m}}^{\text{AB}}-\text{y}_{\text{m}}^{\text{AE}}\right)+\sqrt{\chi _+} \phi _+ \left(\text{y}_{\text{m}}^{\text{EE}}-\text{y}_{\text{m}}^{\text{BB}}\right)\right]
\]
\[
\left\langle \Gamma _{\text{2,m}}^{24} \right\rangle=\sqrt{\chi _- \chi _+} \left(\text{y}_{\text{m}}^{\text{BB}}-2 \text{y}_{\text{m}}^{\text{BE}}-\text{y}_{\text{m}}^{\text{CC}}+2 \text{y}_{\text{m}}^{\text{CD}}-\text{y}_{\text{m}}^{\text{DD}}+\text{y}_{\text{m}}^{\text{EE}}\right)+\chi _- \left(\text{y}_{\text{m}}^{\text{BC}}-\text{y}_{\text{m}}^{\text{BD}}-\text{y}_{\text{m}}^{\text{CE}}+\text{y}_{\text{m}}^{\text{DE}}\right)+\chi _+ \left(-\text{y}_{\text{m}}^{\text{BC}}+\text{y}_{\text{m}}^{\text{BD}}+\text{y}_{\text{m}}^{\text{CE}}-\text{y}_{\text{m}}^{\text{DE}}\right)
\]
\[
\left\langle \Gamma _{\text{2,m}}^{33} \right\rangle=\frac{1}{16} \left[2 \phi _- \phi _+ \left(\text{y}_{\text{m}}^{\text{BC}}+\text{y}_{\text{m}}^{\text{BD}}+\text{y}_{\text{m}}^{\text{CE}}+\text{y}_{\text{m}}^{\text{DE}}\right)-8 \phi _- \left(\text{y}_{\text{m}}^{\text{AC}}+\text{y}_{\text{m}}^{\text{AD}}\right)+\phi _-^2 \left(\text{y}_{\text{m}}^{\text{CC}}+2 \text{y}_{\text{m}}^{\text{CD}}+\text{y}_{\text{m}}^{\text{DD}}\right)-8 \phi _+ \left(\text{y}_{\text{m}}^{\text{AB}}+\text{y}_{\text{m}}^{\text{AE}}\right)+\phi _+^2 \left(\text{y}_{\text{m}}^{\text{BB}}+2 \text{y}_{\text{m}}^{\text{BE}}+\text{y}_{\text{m}}^{\text{EE}}\right)+16 \text{y}_{\text{m}}^{\text{AA}}\right]
\]
\[
\left\langle \Gamma _{\text{2,m}}^{34} \right\rangle=\frac{1}{4} \left[\sqrt{\chi _-} \phi _+ \left(\text{y}_{\text{m}}^{\text{EE}}-\text{y}_{\text{m}}^{\text{BB}}\right)+\phi _- \sqrt{\chi _+} \left(\text{y}_{\text{m}}^{\text{CC}}-\text{y}_{\text{m}}^{\text{DD}}\right)+4 \sqrt{\chi _-} \left(\text{y}_{\text{m}}^{\text{AB}}-\text{y}_{\text{m}}^{\text{AE}}\right)+\sqrt{\chi _-} \phi _- \left(-\text{y}_{\text{m}}^{\text{BC}}-\text{y}_{\text{m}}^{\text{BD}}+\text{y}_{\text{m}}^{\text{CE}}+\text{y}_{\text{m}}^{\text{DE}}\right)-4 \sqrt{\chi _+} \left(\text{y}_{\text{m}}^{\text{AC}}-\text{y}_{\text{m}}^{\text{AD}}\right)+\sqrt{\chi _+} \phi _+ \left(\text{y}_{\text{m}}^{\text{BC}}-\text{y}_{\text{m}}^{\text{BD}}+\text{y}_{\text{m}}^{\text{CE}}-\text{y}_{\text{m}}^{\text{DE}}\right)\right]
\]
\[
\left\langle \Gamma _{\text{2,m}}^{44} \right\rangle=\sqrt{\chi _- \chi _+} \left(-2 \text{y}_{\text{m}}^{\text{BC}}+2 \text{y}_{\text{m}}^{\text{BD}}+2 \text{y}_{\text{m}}^{\text{CE}}-2 \text{y}_{\text{m}}^{\text{DE}}\right)+\chi _- \left(\text{y}_{\text{m}}^{\text{BB}}-2 \text{y}_{\text{m}}^{\text{BE}}+\text{y}_{\text{m}}^{\text{EE}}\right)+\chi _+ \left(\text{y}_{\text{m}}^{\text{CC}}-2 \text{y}_{\text{m}}^{\text{CD}}+\text{y}_{\text{m}}^{\text{DD}}\right)
\]
\label{eq:10_component_pair_occupation}
\end{subequations}
\end{widetext}

Rewriting Eq.(\ref{eq:CE}) in term of average point and pair cluster functions given by Eqs.(7) and (8), it is found that the configurational enthalpy of mixing for quinary alloys can be expressed as a function of concentration $c_{p}$ and average pair probabilities $\text{y}_m^{\text{pq}}$ by the following expression:

\begin{widetext}
	\[
	\text{$\Delta $H}_{\text{mix}} ( \vec{\sigma } )=J_{1,1}^0 \left\langle \Gamma _{1,1}^0( \vec{\sigma } ) \right\rangle +J_{1,1}^1 \left\langle \Gamma _{1,1}^1( \vec{\sigma } ) \right\rangle +J_{1,1}^2  \left\langle \Gamma _{1,1}^2( \vec{\sigma } ) \right\rangle 
	+J_{1,1}^3 \left\langle \Gamma _{1,1}^3( \vec{\sigma } ) \right\rangle +
	+J_{1,1}^4 \left\langle \Gamma _{1,1}^4( \vec{\sigma } ) \right\rangle +
	\sum _{\text{m pairs}} \left[m_{\text{2,m}}^{11} J_{\text{2,m}}^{11}  \left\langle \Gamma _{\text{2,m}}^{11}( \vec{\sigma } ) \right\rangle +m_{\text{2,m}}^{12} J_{\text{2,m}}^{12}  \left\langle \Gamma _{\text{2,m}}^{12}( \vec{\sigma } ) \right\rangle +m_{\text{2,m}}^{13} J_{\text{2,m}}^{13}  \left\langle \Gamma _{\text{2,m}}^{13}( \vec{\sigma } ) \right\rangle +m_{\text{2,m}}^{14} J_{\text{2,m}}^{14}  \left\langle \Gamma _{\text{2,m}}^{14}( \vec{\sigma } ) \right\rangle +m_{\text{2,m}}^{22} J_{\text{2,m}}^{22} \left\langle \Gamma _{\text{2,m}}^{22}( \vec{\sigma } ) \right\rangle +m_{\text{2,m}}^{23} J_{\text{2,m}}^{23}  \left\langle \Gamma _{\text{2,m}}^{23}( \vec{\sigma } ) \right\rangle +m_{\text{2,m}}^{24} J_{\text{2,m}}^{24} \left\langle \Gamma _{\text{2,m}}^{24}( \vec{\sigma } ) \right\rangle +m_{\text{2,m}}^{33} J_{\text{2,m}}^{33} \left\langle \Gamma _{\text{2,m}}^{33}( \vec{\sigma } ) \right\rangle \text{}+m_{\text{2,m}}^{34} J_{\text{2,m}}^{34} \left\langle \Gamma _{\text{2,m}}^{34}( \vec{\sigma } ) \right\rangle \text{}+m_{\text{2,m}}^{44} J_{\text{2,m}}^{44}  \left\langle \Gamma _{\text{2,m}}^{44}( \vec{\sigma } ) \right\rangle  \text{}\right]+\sum _{\text{triplets}}\cdots
	\]
\end{widetext}

The ECI $J_{\omega }$ parameters in five-component alloy MoNbTaVW system were obtained by mapping DFT energies calculated for 428 bcc-like structures from different binaries, ternaries, quaternaries into the CE Hamiltonian in Eq.(9) by using the structure inversion method (SIM) [\onlinecite{Sanchez1984,Connolly1983}]. The fitting procedure was carried out using the ATAT package [\onlinecite{vandeWalle2002}] and the cross-validation error between DFT and CE energies was about 8 meV/atom. The values of 30 pair and 40 triple ECIs were reported in 
[\onlinecite{Toda-Caraballo2015}]. It is found that the dominant contributions to the pairwise energies come from the first and second bcc nearest neighbor interactions whereas the third nearest-neighbor pair interactions and the three-body interactions are significantly smaller. It is worth noting that the CE energy for multi-component alloy in Eq.(9) represents a generalization of the Ising-like Hamiltonian where the only nearest neighbor interactions between different species are taken into account, for example, in the case of four-component HEAs MoNbTaW [\onlinecite{Widom2013}]. The CE Hamiltonian can be used to perform quasi-canonical Monte-Carlo simulations in order to investigate free energies of alloy formation from disorder to order phase transitions. The evaluations of the free energies involved thermodynamic integration algorithm for computing the configurational entropies of alloys as it has been described in detail previously [\onlinecite{Wrobel2015}]. More importantly, the chemical short-range order (SRO) parameters describing the occupational derivations from the average random configuration at a local atomic scale can be evaluated from the CE free energies of mixing and compared with available experimental data.    

\begin{widetext}
\section{Short range order parameters for multicomponent alloy}

The degree of chemical SRO influences both the configuration entropy and enthalpy of mixing in alloy complex. The chemical SRO is absent in ideal solid solutions where atom species occupy sites randomly. It is common to describe SRO in form of Warren-Cowley short-range order or pair-correlation parameters in the following formula [\onlinecite{deFontaine1971}]: 
\[
\alpha _{\text{2,m}}^{\text{pq}}=1-\frac{\text{y}_{m}^{\text{pq}}}{\text{c}_{\text{p}} \text{c}_{\text{q}}}
\label{eq:Warren-Cowley-SRO-definition}
\]

Here we use the notation for the average concentration $c_{p}$ and the average pair probability for the $m^{th}$ nearest-neighbor shell $y_{m}^{pq}$ as they have been defined from Eqs.(5) and (6), respectively. Note that $P_{m}^{pq}$=$y_{m}^{pq}$/$c_{p}$ is the conditional probability of finding atom q in the $m^{th}$ coordination shell surrounding the atom p. When $\alpha_{\text{2,m}}^{\text{pq}}=0$ this describes random alloys namely in this case elements p and q in the pair configuration m are found in the alloy system with a probability equal to $\text{c}_{\text{p}}\text{c}_{\text{q}}$. In the case of $\alpha_{\text{2,m}}^{\text{pq}}>0$ there is a tendency of clustering or segregation of p-p and q-q pairs and for	$\alpha_{\text{2,m}}^{\text{pq}}<0$ there is a tendency of unlike pairs ordering p-q.

Short ranger order parameters can be expressed in terms of average point and pair correlation functions. The expressions for the $\text{y}_{2,m}^{\text{pq}}$ in Eq.\ref{eq:Warren-Cowley-SRO-definition} can be obtained from the inversion of the equations \ref{eq:10_component_pair_occupation} and \ref{eq:5_component_point_corr_functions}. For a five component alloy system there are 10 distinct pair probability functions and their explicit formulas are below. 

\begin{subequations}
\[
y _{\text{2,m}}^{\text{AB}}=\frac{1}{50} \left[-(\kappa +3) \left\langle \Gamma _{\text{1,1}}^1 \right\rangle +\sqrt{2} \sqrt{\kappa +5} \left(2 \left( \left\langle \Gamma _{\text{2,m}}^{12} \right\rangle +\left\langle \Gamma _{\text{2,m}}^{23} \right\rangle \right)- \left\langle \Gamma _{\text{1,1}}^2 \right\rangle \right)+\sqrt{10-2 \kappa } \left(2 \left( \left\langle \Gamma _{\text{2,m}}^{14} \right\rangle + \left\langle \Gamma _{\text{2,m}}^{34} \right\rangle \right)- \left\langle \Gamma _{\text{1,1}}^4 \right\rangle \right)+(\kappa -3) \left\langle \Gamma _{\text{1,1}}^3 \right\rangle +2 (\kappa -1) \left\langle \Gamma _{\text{2,m}}^{11} \right\rangle -4 \left\langle \Gamma _{\text{2,m}}^{13} \right\rangle-2 (\kappa +1) \left\langle \Gamma _{\text{2,m}}^{33} \right\rangle +2\right]
\]
\[
y _{\text{2,m}}^{\text{AC}}=\frac{1}{50} \left[(\kappa -3) \left\langle \Gamma _{\text{1,1}}^1 \right\rangle +\sqrt{10-2 \kappa } \left(2 \left( \left\langle \Gamma _{\text{2,m}}^{12} \right\rangle + \left\langle \Gamma _{\text{2,m}}^{23} \right\rangle \right)- \left\langle \Gamma _{\text{1,1}}^2 \right\rangle \right)+\sqrt{2} \sqrt{\kappa +5} \left( \left\langle \Gamma _{\text{1,1}}^4 \right\rangle -2 \left(\left\langle \Gamma _{\text{2,m}}^{14} \right\rangle + \left\langle \Gamma _{\text{2,m}}^{34} \right\rangle \right)\right)-(\kappa +3) \left\langle \Gamma _{\text{1,1}}^3 \right\rangle -2 (\kappa +1) \left\langle \Gamma _{\text{2,m}}^{11} \right\rangle -4 \left\langle \Gamma _{\text{2,m}}^{13} \right\rangle +2 (\kappa -1) \left\langle \Gamma _{\text{2,m}}^{33} \right\rangle +2\right]
\]
\[
y _{\text{2,m}}^{\text{AD}}=\frac{1}{50} \left[(\kappa -3) \left\langle \Gamma _{\text{1,1}}^1 \right\rangle +\sqrt{10-2 \kappa } \left( \left\langle \Gamma _{\text{1,1}}^2 \right\rangle -2 \left( \left\langle \Gamma _{\text{2,m}}^{12} \right\rangle + \left\langle \Gamma _{\text{2,m}}^{23} \right\rangle \right)\right)+\sqrt{2} \sqrt{\kappa +5} \left(2 \left( \left\langle \Gamma _{\text{2,m}}^{14} \right\rangle + \left\langle \Gamma _{\text{2,m}}^{34} \right\rangle \right)- \left\langle \Gamma _{\text{1,1}}^4 \right\rangle \right)-(\kappa +3) \left\langle \Gamma _{\text{1,1}}^3 \right\rangle -2 (\kappa +1) \left\langle \Gamma _{\text{2,m}}^{11} \right\rangle -4 \left\langle \Gamma _{\text{2,m}}^{13} \right\rangle +2 (\kappa -1) \left\langle \Gamma _{\text{2,m}}^{33} \right\rangle +2\right]
\]
\[
y _{\text{2,m}}^{\text{AE}}=\frac{1}{50} \left[-(\kappa +3) \left\langle \Gamma _{\text{1,1}}^1 \right\rangle +\sqrt{2} \sqrt{\kappa +5} \left(\left\langle \Gamma _{\text{1,1}}^2 \right\rangle -2 \left( \left\langle \Gamma _{\text{2,m}}^{12} \right\rangle + \left\langle \Gamma _{\text{2,m}}^{23} \right\rangle \right)\right)+\sqrt{10-2 \kappa } \left( \left\langle \Gamma _{\text{1,1}}^4 \right\rangle -2 \left( \left\langle \Gamma _{\text{2,m}}^{14} \right\rangle + \left\langle \Gamma _{\text{2,m}}^{34} \right\rangle \right)\right)+(\kappa -3) \left\langle \Gamma _{\text{1,1}}^3 \right\rangle +2 (\kappa -1) \left\langle \Gamma _{\text{2,m}}^{11} \right\rangle -4 \left\langle \Gamma _{\text{2,m}}^{13} \right\rangle -2 (\kappa +1) \left\langle \Gamma _{\text{2,m}}^{33} \right\rangle +2\right]
\]
\[
y _{\text{2,m}}^{\text{BC}}=\frac{1}{25} \left[ \left\langle \Gamma _{\text{1,1}}^1 \right\rangle -\sqrt{2 \kappa +5} \left( \left\langle \Gamma _{\text{1,1}}^2 \right\rangle + \left\langle \Gamma _{\text{2,m}}^{14} \right\rangle \right)+\sqrt{5-2 \kappa } \left( \left\langle \Gamma _{\text{1,1}}^4 \right\rangle - \left\langle \Gamma _{\text{2,m}}^{23} \right\rangle \right)+ \left\langle \Gamma _{\text{1,1}}^3 \right\rangle - \left\langle \Gamma _{\text{2,m}}^{11} \right\rangle -\sqrt{10-2 \kappa } \left\langle \Gamma _{\text{2,m}}^{12} \right\rangle +3 \left\langle \Gamma _{\text{2,m}}^{13} \right\rangle +\kappa  \left\langle \Gamma _{\text{2,m}}^{22} \right\rangle -\kappa  \left\langle \Gamma _{\text{2,m}}^{24} \right\rangle- \left\langle \Gamma _{\text{2,m}}^{33} \right\rangle +\sqrt{2} \sqrt{\kappa +5} \left\langle \Gamma _{\text{2,m}}^{34} \right\rangle -\kappa  \left\langle \Gamma _{\text{2,m}}^{44} \right\rangle +1\right]
\]
\[
y _{\text{2,m}}^{\text{BD}}=\frac{1}{100} \left[4 \left\langle \Gamma _{\text{1,1}}^1 \right\rangle -4 \sqrt{5-2 \kappa } \left\langle \Gamma _{\text{1,1}}^2 \right\rangle +4 \left\langle \Gamma _{\text{1,1}}^3 \right\rangle -4 \sqrt{2 \kappa +5} \left\langle \Gamma _{\text{1,1}}^4 \right\rangle -4 \left\langle \Gamma _{\text{2,m}}^{11} \right\rangle +\sqrt{2} \sqrt{\kappa +5} \left(-4 \left\langle \Gamma _{\text{2,m}}^{12} \right\rangle +(\kappa -3) \left\langle \Gamma _{\text{2,m}}^{14} \right\rangle +(\kappa +1) \left\langle \Gamma _{\text{2,m}}^{23} \right\rangle \right)+4 \left(3 \left\langle \Gamma _{\text{2,m}}^{13} \right\rangle -\kappa  \left\langle \Gamma _{\text{2,m}}^{22} \right\rangle +\kappa  \left\langle \Gamma _{\text{2,m}}^{24} \right\rangle -\left\langle \Gamma _{\text{2,m}}^{33} \right\rangle -\sqrt{10-2 \kappa } \left\langle \Gamma _{\text{2,m}}^{34} \right\rangle +\kappa  \left\langle \Gamma _{\text{2,m}}^{44} \right\rangle \right)+4\right]
\]
\[
y _{\text{2,m}}^{\text{BE}}=\frac{1}{50} \left[-2 (\kappa -1) \left\langle \Gamma _{\text{1,1}}^1 \right\rangle +2 (\kappa +1) \left\langle \Gamma _{\text{1,1}}^3 \right\rangle -(\kappa -3) \left\langle \Gamma _{\text{2,m}}^{11} \right\rangle -4 \left\langle \Gamma _{\text{2,m}}^{13} \right\rangle -(\kappa +5) \left\langle \Gamma _{\text{2,m}}^{22} \right\rangle -4 \kappa  \left\langle \Gamma _{\text{2,m}}^{24} \right\rangle+(\kappa +3) \left\langle \Gamma _{\text{2,m}}^{33} \right\rangle +(\kappa -5) \left\langle \Gamma _{\text{2,m}}^{44} \right\rangle +2\right]
\]
\[
y _{\text{2,m}}^{\text{CD}}=\frac{1}{50} \left[2 (\kappa +1) \left\langle \Gamma _{\text{1,1}}^1 \right\rangle -2 (\kappa -1) \left\langle \Gamma _{\text{1,1}}^3 \right\rangle +(\kappa +3) \left\langle \Gamma _{\text{2,m}}^{11} \right\rangle -4 \left\langle \Gamma _{\text{2,m}}^{13} \right\rangle +(\kappa -5) \left\langle \Gamma _{\text{2,m}}^{22} \right\rangle +4 \kappa  \left\langle \Gamma _{\text{2,m}}^{24} \right\rangle -(\kappa -3) \left\langle \Gamma _{\text{2,m}}^{33} \right\rangle -(\kappa +5) \left\langle \Gamma _{\text{2,m}}^{44} \right\rangle +2\right]
\]
\[
y _{\text{2,m}}^{\text{CE}}=\frac{1}{100} \left[4 \left\langle \Gamma _{\text{1,1}}^1 \right\rangle +4 \sqrt{5-2 \kappa } \left\langle \Gamma _{\text{1,1}}^2 \right\rangle +4 \left\langle \Gamma _{\text{1,1}}^3 \right\rangle +4 \sqrt{2 \kappa +5} \left\langle \Gamma _{\text{1,1}}^4 \right\rangle -4 \left\langle \Gamma _{\text{2,m}}^{11} \right\rangle -\sqrt{2} \sqrt{\kappa +5} \left(-4 \left\langle \Gamma _{\text{2,m}}^{12} \right\rangle +(\kappa -3) \left\langle \Gamma _{\text{2,m}}^{14} \right\rangle +(\kappa +1) \left\langle \Gamma _{\text{2,m}}^{23} \right\rangle \right)+4 \left(3 \left\langle \Gamma _{\text{2,m}}^{13} \right\rangle -\kappa  \left\langle \Gamma _{\text{2,m}}^{22} \right\rangle +\kappa  \left\langle \Gamma _{\text{2,m}}^{24} \right\rangle -\left\langle \Gamma _{\text{2,m}}^{33} \right\rangle +\sqrt{10-2 \kappa } \left\langle \Gamma _{\text{2,m}}^{34} \right\rangle +\kappa  \left\langle \Gamma _{\text{2,m}}^{44} \right\rangle \right)+4\right]
\]
\[
y _{\text{2,m}}^{\text{DE}}=\frac{1}{25} \left[\left\langle \Gamma _{\text{1,1}}^1 \right\rangle +\sqrt{2 \kappa +5} \left( \left\langle \Gamma _{\text{1,1}}^2 \right\rangle +\left\langle \Gamma _{\text{2,m}}^{14}\right\rangle \right)+\sqrt{5-2 \kappa } \left( \left\langle \Gamma _{\text{2,m}}^{23} \right\rangle -\left\langle \Gamma _{\text{1,1}}^4 \right\rangle \right)+ \left\langle \Gamma _{\text{1,1}}^3 \right\rangle -\left\langle \Gamma _{\text{2,m}}^{11} \right\rangle +\sqrt{10-2 \kappa } \left\langle \Gamma _{\text{2,m}}^{12} \right\rangle +3 \left\langle \Gamma _{\text{2,m}}^{13} \right\rangle +\kappa  \left\langle \Gamma _{\text{2,m}}^{22} \right\rangle -\kappa  \left\langle \Gamma _{\text{2,m}}^{24} \right\rangle -\left\langle \Gamma _{\text{2,m}}^{33} \right\rangle -\sqrt{2} \sqrt{\kappa +5} \left\langle \Gamma _{\text{2,m}}^{34} \right\rangle -\kappa  \left\langle \Gamma _{\text{2,m}}^{44} \right\rangle +1\right]
\]
\label{eq:2-point-occupation-probabilities}
\end{subequations}
\end{widetext}

where $\kappa=\sqrt{5}$. Substituting Eqs.(11) into (10), the full set of ten different chemical SRO parameters $\alpha_{2,m}^{pq}$ can be calculated using both point and pair correlation functions generated by the Monte-Carlo simulations with the ECIs. The numerical value of each of the cluster correlation functions are obtained from the cluster expansion of the Hamiltonian in combination with Monte Carlo simulations as a function of temperature. The resulting Warren-Cowley SRO parameters : $\alpha_{\text{2,m}}^{\text{AB}}$, 
$\alpha_{\text{2,m}}^{\text{AC}}$,...,$\alpha_{\text{2,m}}^{\text{ED}}$ as a function of alloy temperature for equimolar HEAs Mo-Nb-Ta-V-W will be analyzed in the next section.

\section{Application high entropy alloy Mo-Nb-Ta-V-W}

\begin{figure}[H]
	\includegraphics[width=\linewidth]{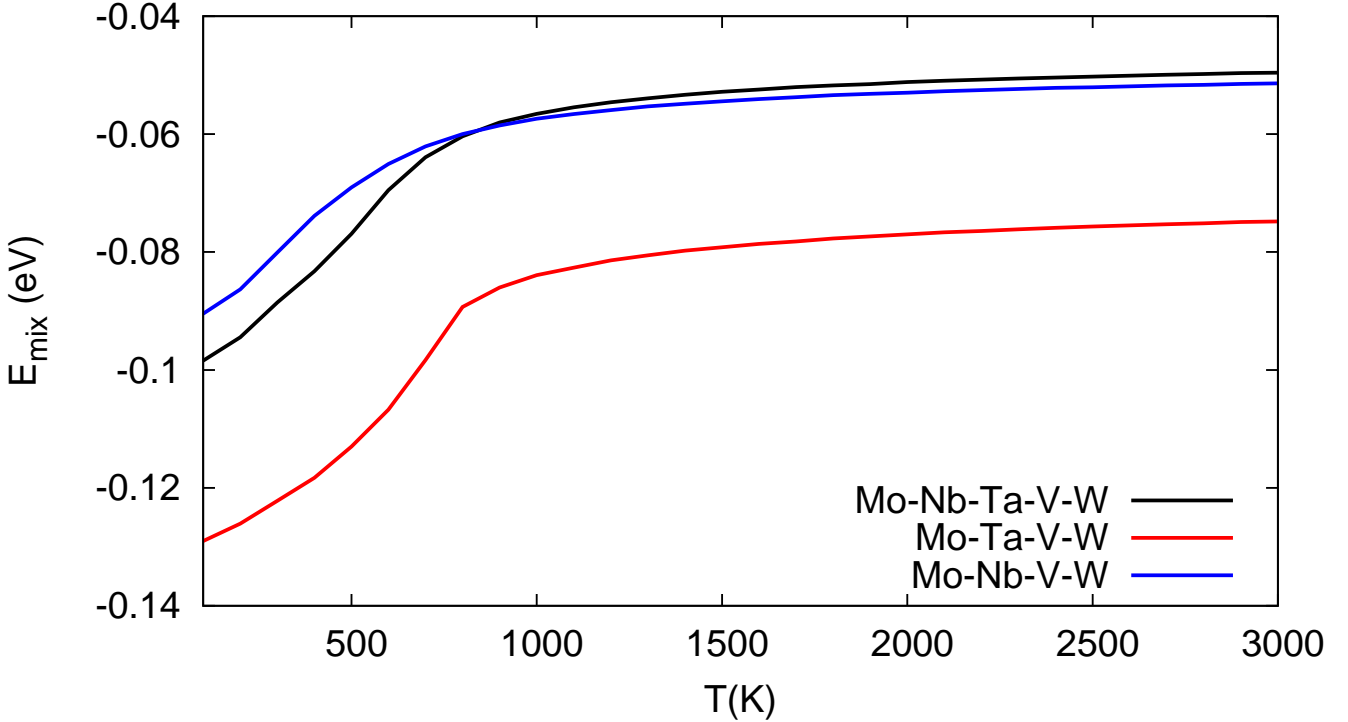}	
	\caption{
		Enthalpy of mixing of the three equimolar HEAs: quinary Mo-Nb-Ta-V-W and quaternaries Mo-Ta-V-W, Mo-Nb-V-W, as a function of temperature.}
	\label{fig:enthalpy_mix_temperature}
\end{figure}

Using the effective cluster interactions (ECI) computed for quinary system Mo-Nb-Ta-V-W [\onlinecite{Toda-Caraballo2015}] the energetically favorable atomic configurations are investigated as a function temperature and alloy composition using quasi-canonical Monte-Carlo simulations. Fig. \ref{eq:basic_enthalpy_mix} shows the evolution of mixing enthalpy as a function of temperature from 3000 K for the equimolar quinary Mo-Nb-Ta-V-W system as well as for the two quaternary sub-systems: Mo-Ta-V-W and W-Nb-V-W. These pre-melting configurations were generated by random numbers for the case of large systems whereas for the case of smaller simulation cell special quasi-random structures (SQSs) can be used [\onlinecite{Zunger1990}]. Inflection points from the temperature curve of enthalpy of mixing indicate order-disorder phase transformations from solid solution phase into different non-random configurations. From Fig.\ref{eq:basic_enthalpy_mix}, at temperatures below 750 K, a partially ordered phase is thermodynamically more stable than the equimolar solid-solution random configuration. For the two sub-system alloys, it is found that the enthalpy of mixing for HEAs in the presence of Mo-Ta binaries, namely Mo-Ta-V-W, is significantly lower not only than the those for the Mo-Nb-V-W where Ta element is absent, but also in a comparison with quinary Mo-Nb-Ta-V-W system.

\begin{figure}[H]
	\includegraphics[width=\linewidth]{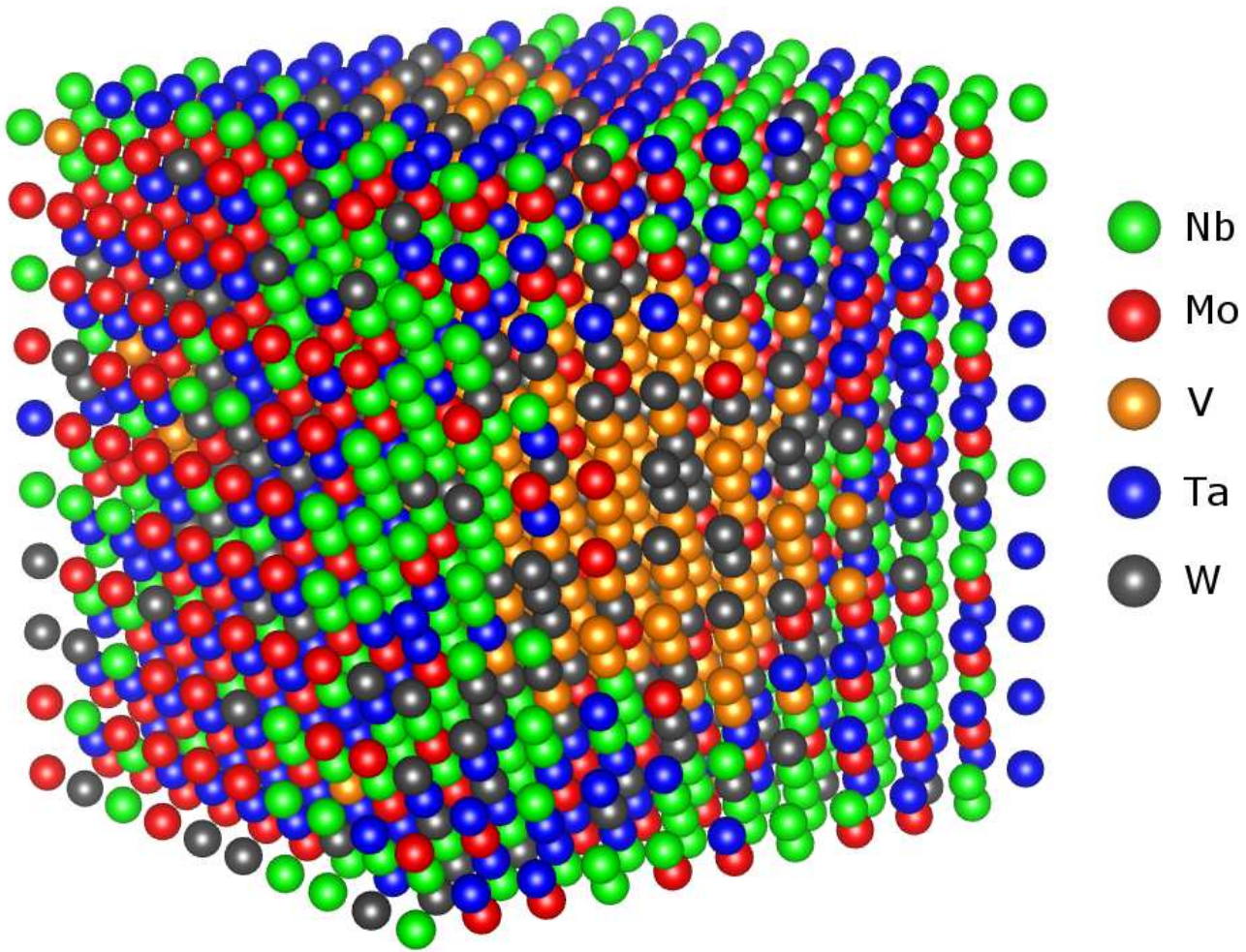}	
	\caption{
		Atomistic configuration for equimolar Mo-Nb-V-Ta-W HEAs obtained from the present MC simulations at 400 K}
	\label{fig:atomistic_picture_400K}
\end{figure}

The simulated structure of this alloy system is generated at T=400K and depicted in Fig.\ref{fig:atomistic_picture_400K}. Phase segregation or clustering of Nb (in green) around the edges of the atomic cell and of V (in yellow) around the centre of the cell can be seen clearly from the presented configuration implying that there are no chemical attractive interactions between Nb and V atoms. The latter finding is consistent with our DFT calculations for bcc Nb-V alloy that the enthalpies of mixing for this binary are positive in all the composition range from our DFT calculations. On the contrary the chemical ordering of Mo-Ta (red and blue) can be appreciated on the left hand side face of the cubic cell that is in a full agreement with the strong negative enthalpy of mixing for the bcc binary Mo-Ta system. More detailed analysis of chemical SRO parameters will confirm the above observations. 

\begin{figure}[H]
	\includegraphics[width=\linewidth]{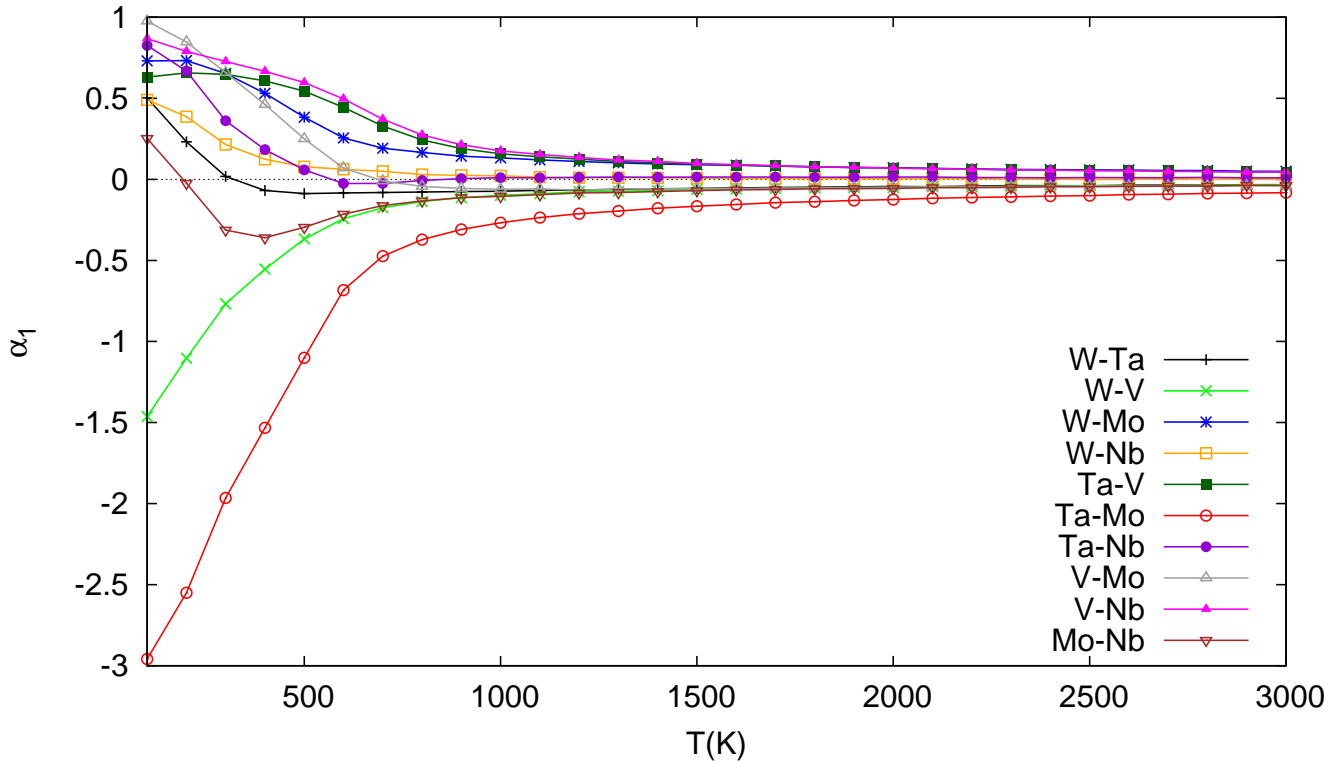}	
	\caption{
		Evolution of 1NN short-range-order parameters in equimolar HEAs Mo-Nb-Ta-V-W as a function temperature.}
	\label{fig:SRO_all}
\end{figure}

For the five-component alloy, the SRO formulation presented in previous section can be applied to study order-disorder trends of the high-entropy Mo-Nb-Ta-V-W system  as a function composition and temperature. The temperature-dependent evolution of ten different SRO parameters for the first nearest-neighbor (NN) shell for the equimolar composition is shown in Fig.\ref{fig:SRO_all}. In the high-temperature limit, it can be seen clearly from Fig.\ref{fig:SRO_all} that all the SRO parameters trend to zero value corresponding to the presence of ideal randomly single phase of solid solution with the configuration entropy of mixing $\Delta S_{mix}$=-R$\sum_{p}c_{p}lnc_{p}$ (where R is the gas constant). The latter expression is conventionally used in the definition of HEAs [\onlinecite{Zhang2014}]. In the temperature region lower than 750K,  from Fig.\ref{fig:SRO_all} it is found that the SRO parameter for Mo-Ta pair, $\alpha_{\text{2,1}}^{\text{Mo-Ta}}$,  becomes the most negative one. This indicates a strong probability of having Ta atoms around a Mo atom in the first shell of the considered equimolar Mo-Ta-Ta-V-W bcc alloy. The prediction of strong chemical SRO parameter for Mo-Ta pair is in a consistent agreement with the previous DFT study for the first nearest-neighbor Mo-Ta bonding interaction in the quaternary equimolar Mo-Nb-Ta-W system [\onlinecite{Widom2013}] and the present study of DFT data-base for the negative enthalpy of mixing in bcc binary Mo-Ta system. The next and strong negative SRO parameter has been predicted for V-W pair from our MC simulation and it is followed by the SRO parameter for Mo-Nb pair. Note that the negative SRO parameters for Mo-Ta, V-W and Mo-Nb are in agreement with the energetically phase stability trend of bcc binary systems between groups V and VI from the periodic table of elements [\onlinecite{Blum2005}]. It is interesting to find from Fig.\ref{fig:SRO_all} that the SRO parameter for Mo-Nb pair becomes positive at very low temperature showing that there is competition between Nb and Ta atoms to occupy the first nearest-neighbor shell surrounding Mo atom in the five-component alloy. At low temperature region there is strong tendency of segregation with the positive values of $\alpha_{\text{2,1}}^{\text{NbV}}$, $\alpha_{\text{2,1}}^{\text{MoW}}$ and $\alpha_{\text{2,1}}^{\text{TaV}}$ namely for SRO parameters of the pairs between elements of the same groups V or VI. 

Having said that it is worth mentioning here that the behavior of first nearest-neighbor SRO parameter in the multi-component bcc-HEAs is much more complex than those predicted from the related binary system. For example, interpretation of the favorable V-W and unfavorable Ta-W 1NN chemical bonding obtained from Fig.\ref{fig:SRO_all} for the negative $\alpha_{\text{2,1}}^{\text{VW}}$ and positive $\alpha_{\text{2,1}}^{\text{TaW}}$ value, respectively, at very low temperature seems to be quite different to those predicted by the DFT calculations of the ground-state structure of $B32$ for V-W and $B2_{3}$ for Ta-W binary systems at equimolar composition [\onlinecite{Muzyk2011}]. It is known that the 1NN environment in $B32$ structure is not fully favorable for the chemical interaction in a comparison with those in the $B2$ or $B2_{3}$ structure. It is also important to emphasize again that the present study provides atomistic configurations which have been generated from a set of ECIs which include not only pair-wise interactions but also the triple effective cluster-expansion contributions in Eq.(9). This means that the enthalpy of mixing for five-component system Mo-Nb-V-Ta-W and therefore the configuration entropy as well as the free energy of mixing are certainly determined beyond the simple Ising model with first nearest-neighbor interaction parameter from the ten constituent binary systems.  It can be seen from Fig.\ref{fig:SRO_all} that the SRO parameters $\alpha_{\text{2,1}}^{\text{TaW}}$, $\alpha_{\text{2,1}}^{\text{NbW}}$ and $\alpha_{\text{2,1}}^{\text{MoV}}$ are positive whereas the enthalpy of mixing for the corresponding binaries are negative but smaller than those of the dominant Mo-Ta binary alloys. The complex behavior of interplay between the ten SRO parameters as a function of temperature can be explained by the many-body effects of cluster interactions in stabilizing the multi-component HEAs.

\begin{figure}[H]
	\includegraphics[width=\linewidth]{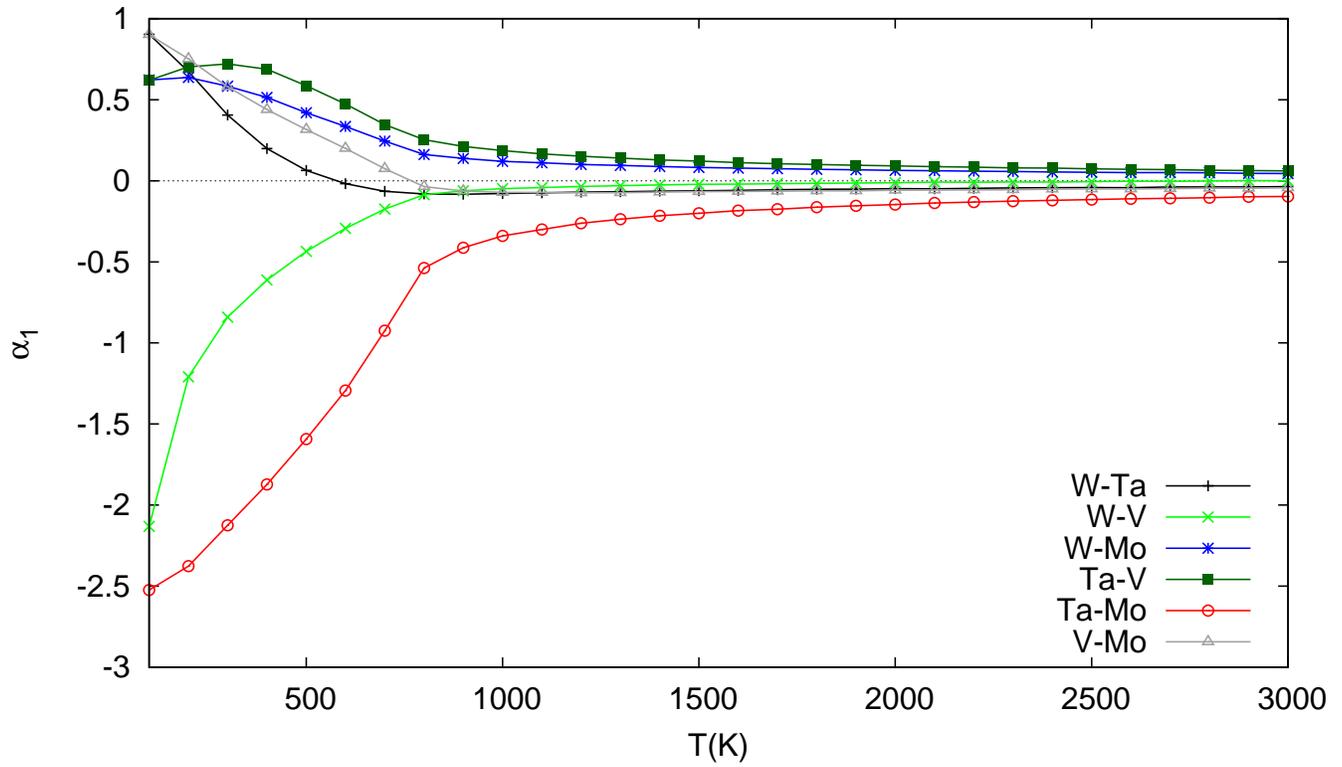}	
	\caption{
		Dependence of 1NN short range order parameters for quaternary subsystem without Nb as a function of temperature.}
	\label{fig:SRO_without_Nb}
\end{figure}

\begin{figure}[H]
	\includegraphics[width=\linewidth]{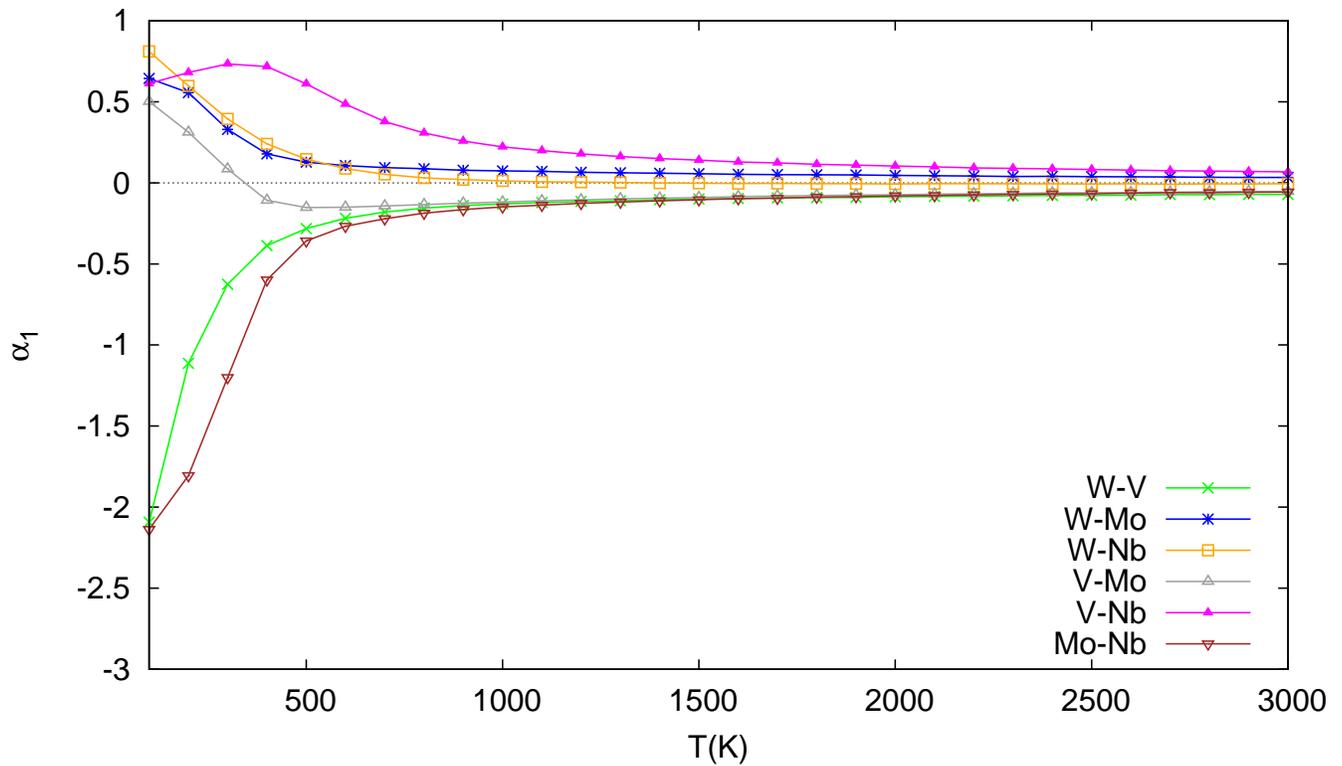}	
	\caption{
		Dependence of 1NN short-range-order parameters for quaternary subsystem without Ta as a function of temperature.}
	\label{fig:SRO_without_Ta}
\end{figure}

In order to cross-check the consistency of SRO expressions for the five component system, in particular the competition between transition metal elements Nb and Ta from group V in the quinary Mo-Nb-Ta-V-W, the SRO parameters for the equimolar and quaternary sub-systems Mo-Ta-V-W and Mo-Nb-V-W are calculated and depicted in Fig.\ref{fig:SRO_without_Nb} and Fig.\ref{fig:SRO_without_Ta}, respectively. Mathematical derivation of the SRO parameters for a quaternary system (A-B-C-D) is carried out by inverting formulas for the correlation functions to get the average pair probabilities (see Eq.\ref{eq:Warren-Cowley-SRO-definition}) in an similar fashion as for the five component system. For a four component system the explicit results in 
$\alpha_{\text{2,m}}^{\text{AB}}$, $\alpha_{\text{2,m}}^{\text{AC}}$,...,$\alpha_{\text{2,m}}^{\text{CD}}$ are as follows. 

\begin{subequations}
\[
\alpha _{\text{2,m}}^{\text{AB}}=1-\frac{-2 \left( \left\langle \Gamma _{\text{1,1}}^1 \right\rangle + \left\langle \Gamma _{\text{1,1}}^2 \right\rangle -2 \left\langle \Gamma _{\text{2,m}}^{12} \right\rangle + \left\langle \Gamma _{\text{2,m}}^{13} \right\rangle - \left\langle \Gamma _{\text{2,m}}^{23} \right\rangle \right)- \left\langle \Gamma _{\text{2,m}}^{33} \right\rangle +1}{\left(-2 \left\langle \Gamma _{\text{1,1}}^1 \right\rangle - \left\langle \Gamma _{\text{1,1}}^3 \right\rangle +1\right) \left(-2 \left\langle \Gamma _{\text{1,1}}^2 \right\rangle + \left\langle \Gamma _{\text{1,1}}^3 \right\rangle +1\right)}
\]	
\[
\alpha _{\text{2,m}}^{\text{AC}}=1-\frac{-2 \left\langle \Gamma _{\text{1,1}}^3 \right\rangle -4 \left\langle \Gamma _{\text{2,m}}^{11} \right\rangle + \left\langle \Gamma _{\text{2,m}}^{33} \right\rangle +1}{\left(1-\left\langle \Gamma _{\text{1,1}}^3 \right\rangle \right){}^2-4 \left( \left\langle \Gamma _{\text{1,1}}^{1} \right\rangle \right)^2}
\]
\[
\alpha _{\text{2,m}}^{\text{AD}}=1-\frac{-2 \left( \left\langle \Gamma _{\text{1,1}}^1 \right\rangle -\left\langle \Gamma _{\text{1,1}}^2 \right\rangle +2 \left\langle \Gamma _{\text{2,m}}^{12} \right\rangle +\left\langle \Gamma _{\text{2,m}}^{13} \right\rangle + \left\langle \Gamma _{\text{2,m}}^{23} \right\rangle \right)- \left\langle \Gamma _{\text{2,m}}^{33} \right\rangle +1}{\left(-2 \left\langle \Gamma _{\text{1,1}}^1 \right\rangle -\left\langle \Gamma _{\text{1,1}}^3 \right\rangle +1\right) \left(2 \left\langle \Gamma _{\text{1,1}}^2 \right\rangle + \left\langle \Gamma _{\text{1,1}}^3 \right\rangle +1\right)}
\]
\[
\alpha _{\text{2,m}}^{\text{BC}}=1-\frac{2 \left( \left\langle \Gamma _{\text{1,1}}^1 \right\rangle -\left\langle \Gamma _{\text{1,1}}^2 \right\rangle -2 \left\langle \Gamma _{\text{2,m}}^{12} \right\rangle +\left\langle \Gamma _{\text{2,m}}^{13} \right\rangle + \left\langle \Gamma _{\text{2,m}}^{23} \right\rangle \right)- \left\langle \Gamma _{\text{2,m}}^{33} \right\rangle +1}{\left(2 \left\langle \Gamma _{\text{1,1}}^1 \right\rangle -\left\langle \Gamma _{\text{1,1}}^3 \right\rangle +1\right) \left(-2 \left\langle \Gamma _{\text{1,1}}^2 \right\rangle +\left\langle \Gamma _{\text{1,1}}^3 \right\rangle +1\right)}
\]
\[
\alpha _{\text{2,m}}^{\text{BD}}=1-\frac{2 \left\langle \Gamma _{\text{1,1}}^3 \right\rangle -4 \left\langle \Gamma _{\text{2,m}}^{22} \right\rangle + \left\langle \Gamma _{\text{2,m}}^{33} \right\rangle +1}{\left(\left\langle \Gamma _{\text{1,1}}^3 \right\rangle +1\right){}^2-4 \left(\left\langle \Gamma _{\text{1,1}}^{2} \right\rangle \right)^2}
\]
\[
\alpha _{\text{2,m}}^{\text{CD}}=1-\frac{2 \left( \left\langle \Gamma _{\text{1,1}}^1 \right\rangle + \left\langle \Gamma _{\text{1,1}}^2 \right\rangle +2 \left\langle \Gamma _{\text{2,m}}^{12} \right\rangle + \left\langle \Gamma _{\text{2,m}}^{13} \right\rangle - \left\langle \Gamma _{\text{2,m}}^{23} \right\rangle \right)- \left\langle \Gamma _{\text{2,m}}^{33} \right\rangle +1}{\left(2 \left\langle \Gamma _{\text{1,1}}^1 \right\rangle - \left\langle \Gamma _{\text{1,1}}^3 \right\rangle +1\right) \left(2 \left\langle \Gamma _{\text{1,1}}^2 \right\rangle + \left\langle \Gamma _{\text{1,1}}^3 \right\rangle +1\right)}
\]	
\end{subequations}

By using Eq.(12), the evolution of six SRO parameters in equimolar Mo-Ta-V-W sub-system is represented in Fig.\ref{fig:SRO_without_Nb}. It is found that the most favorable chemical bonding between the Mo-Ta and V-W pairs in this quaternary sub-system Fig.\ref{fig:SRO_all} are consistent with the results obtained for the corresponding SRO parameters in equimolar Mo-Nb-Ta-V-W system. Here the tendency of phase separation between Mo and W as well as Ta and V in the quaternary Mo-V-Ta-W alloy at the low-temperature region (T $\langle$ 750K) is also demonstrated by positive values of the corresponding parameters $\alpha_{\text{2,1}}^{\text{MoW}}$ and $\alpha_{\text{2,1}}^{\text{TaV}}$. Again the SRO parameters $\alpha_{\text{2,1}}^{\text{TaW}}$, $\alpha_{\text{2,1}}^{\text{MoV}}$ become also positive at the low temperature that is in agreement with the previous discussion for the quinary Mo-Nb-V-Ta-W alloys. Importantly, the opposite feature between 1NN SRO parameters $\alpha_{\text{2,1}}^{\text{VW}}$ and $\alpha_{\text{2,1}}^{\text{TaW}}$ in a comparison with those found in the ground state of $B32$ and $B2_{3}$ structure for VW and TaW binary, respectively, discussed previously in the quinary system remains the same in the quaternary Mo-Ta-V-W.  

In the absence of Ta, the six SRO parameters in the quaternary Mo-Nb-V-W sub-system are represented in Fig.\ref{fig:SRO_without_Ta}. The most interesting result from this figure is that the chemical SRO between Mo and Nb $\alpha_{\text{2,1}}^{\text{MoNb}}$ is now dominantly negative and in a strong competition with $\alpha_{\text{2,1}}^{\text{VW}}$, instead of those for the Mo-Ta pair. This finding indicates the preferred ordering between the Mo-Nb pairs that is a different behavior in a comparison with quinary Mo-Nb-V-Ta-W and quaternary Mo-Ta-V-W alloys where it becomes positive at lower temperatures than 200K.  On the other hand, the SRO for V-W pair is also negative as in previously considered cases for the quinary and the quaternary HEAs. Finally the SRO parameter for Nb-V pair, $\alpha_{\text{2,1}}^{\text{NbV}}$, remains strongly positive for the quaternary Mo-Nb-V-W system in a consistent with its behavior for the initial five-component alloy.

\section{Second nearest-neighbor SRO effects}

\begin{figure}[H]
	\includegraphics[width=\linewidth]{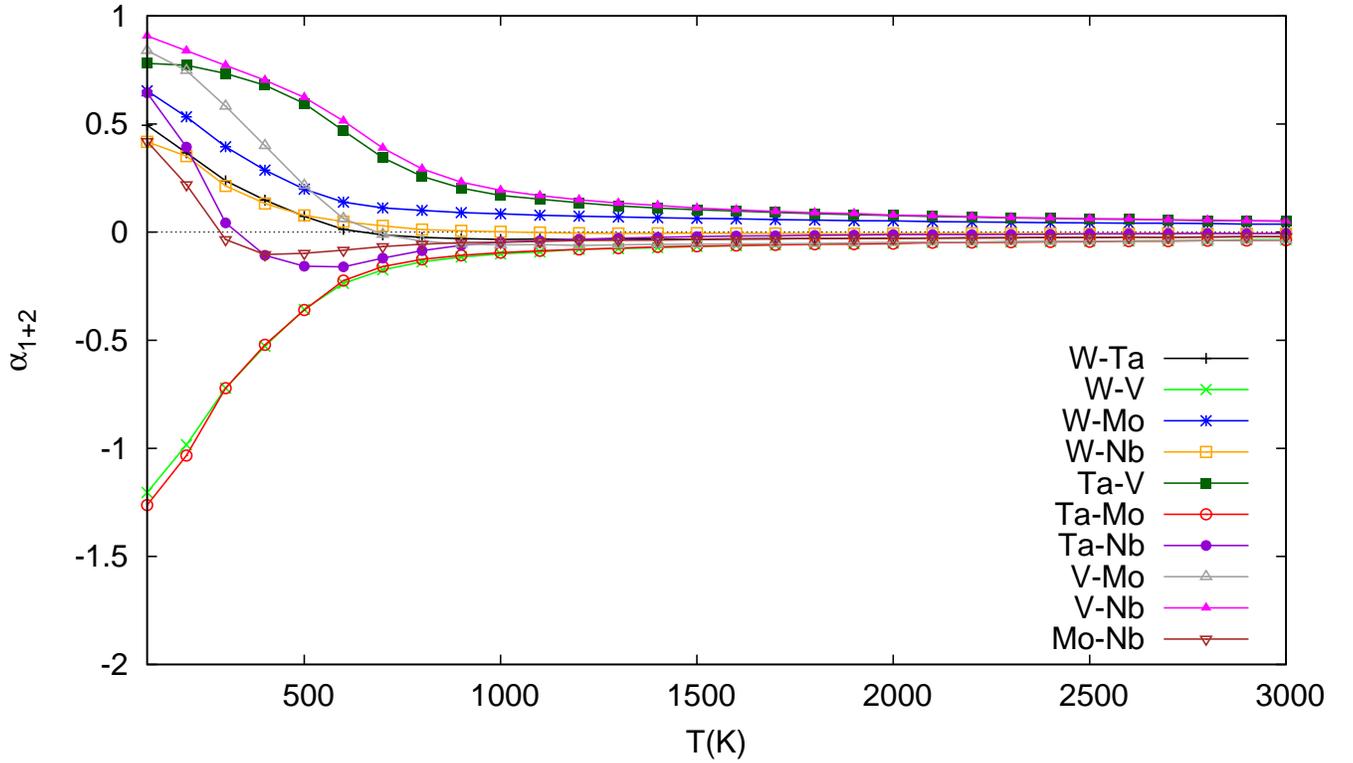}	
	\caption{
		Dependence of average short-range-order parameters for quinary system as a function of temperature.}
	\label{fig:SRO_NNN}
\end{figure}

\begin{figure}[H]
	\includegraphics[width=\linewidth]{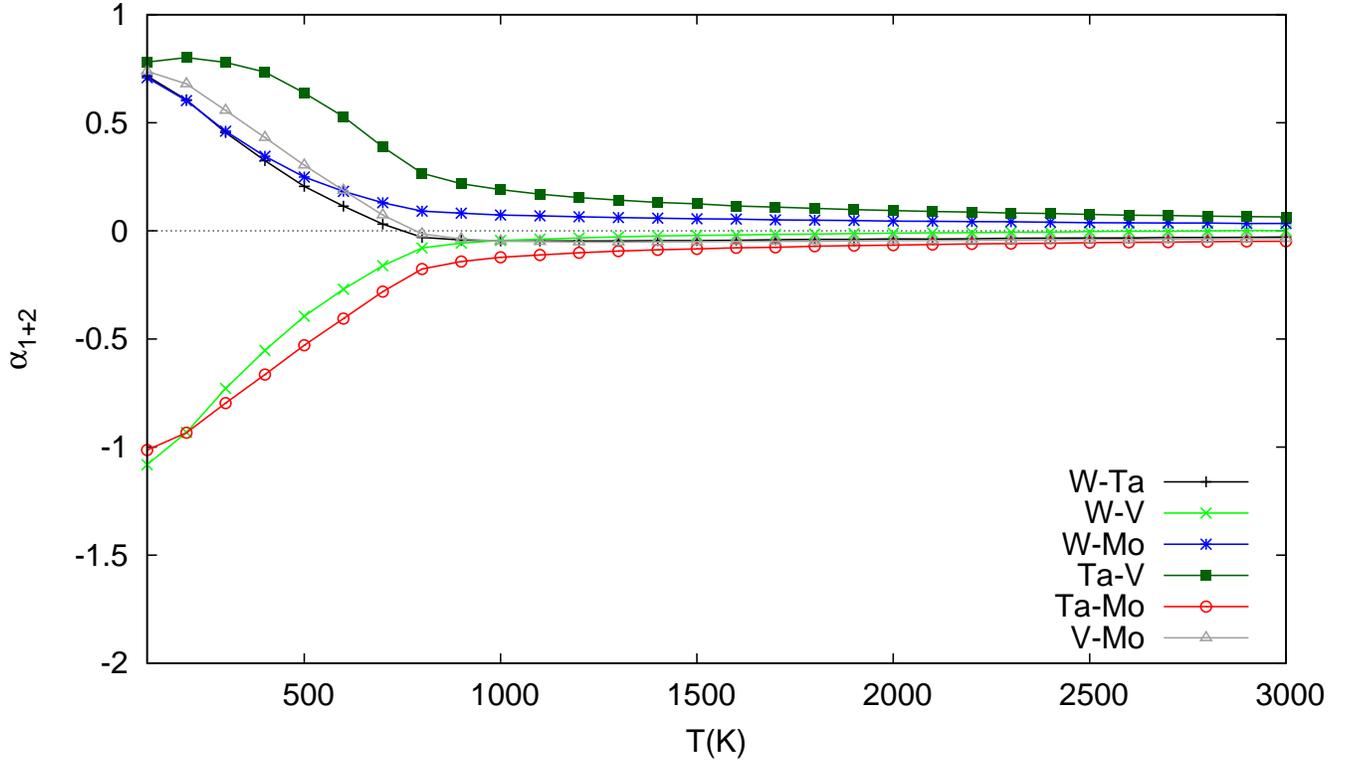}	
	\caption{
		Dependence of average short-range-order parameters for quaternary subsystem without Nb as a function of temperature.}
	\label{fig:SRO_NNN_without_Nb}
\end{figure}

\begin{figure}[H]
	\includegraphics[width=\linewidth]{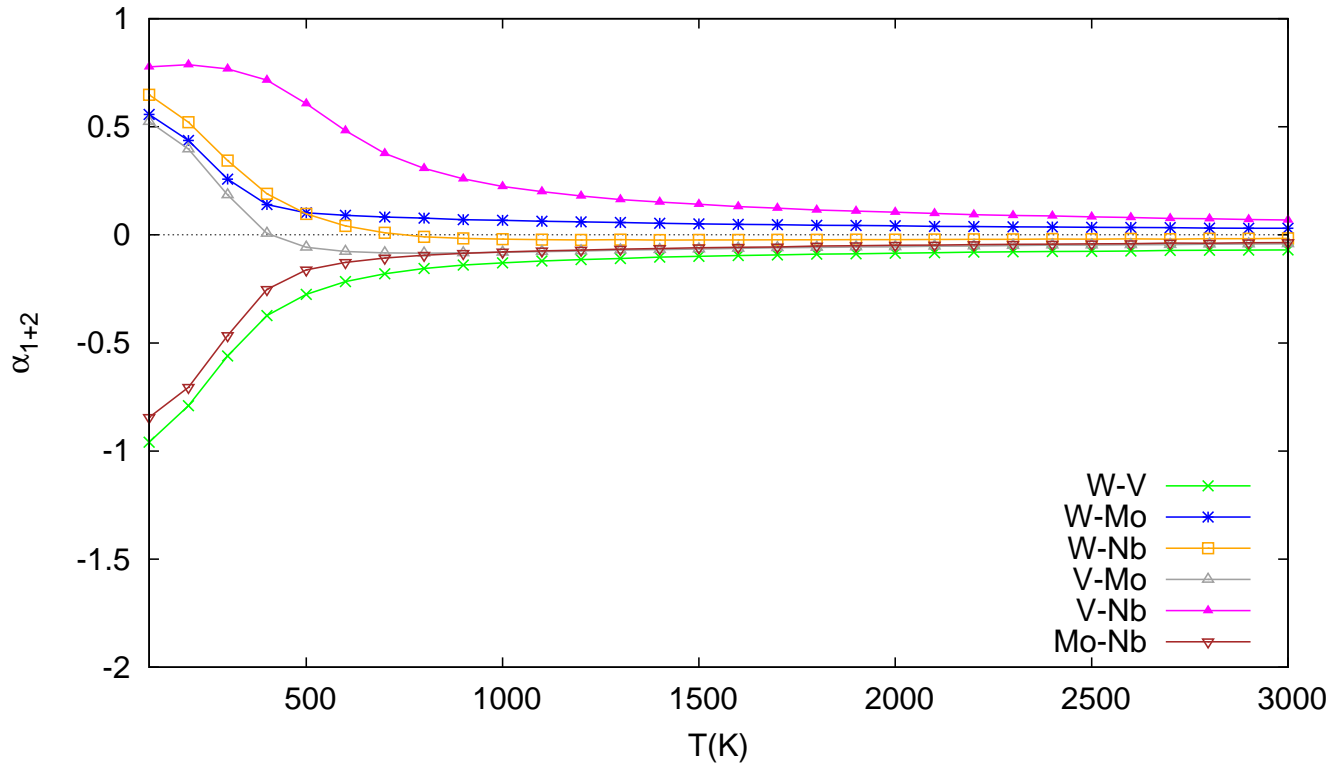}	
	\caption{
		Dependence of average short-range-order parameters for quaternary subsystem without Ta as a function of temperature.}
	\label{fig:SRO_NNN_without_Ta}
\end{figure}

Figs.(\ref{fig:SRO_all}),(\ref{fig:SRO_without_Nb}),(\ref{fig:SRO_without_Ta}) represent SRO parameters in the first nearest-neighbor shell of the three considered HEAs.  For a bcc alloy where the first and second NN distances are very close to each other, the average SRO parameter corresponding to the grouped shell of coordination $z_{1}$+$z_{2}$ is defined as 

\[
\alpha _{\text{1+2}}^{\text{pq}}= \frac{z_{1}\alpha _{\text{2,1}}^{\text{pq}} + z_{2}\alpha _{\text{2,2}}^{\text{pq}}}{z_{1} + z_{2}}
\label{eq:Warren-Cowley-SRO-average}
\]

where $z_{1}$=8 and $z_{2}$=6. Note that the average SRO papameter $\alpha _{\text{1+2}}^{\text{pq}}$ can be deduced experimentally with good confidence from diffuse-neutron scattering measurements [\onlinecite{Mirebeau1984}].  Figs.(\ref{fig:SRO_NNN}),(\ref{fig:SRO_NNN_without_Nb}),(\ref{fig:SRO_NNN_without_Ta}) show the dependence of all average SRO parameters calculated for the equimolar HEAs Mo-Nb-Ta-V-W, Mo-Ta-V-W and Mo-Nb-V-W, respectively. 

Comparing the results obtained from Fig.(\ref{fig:SRO_NNN}) for the average SRO parameters for both 1NN and 2NN, $\alpha _{\text{1+2}}^{\text{pq}}$, with those presented in Fig.(\ref{fig:SRO_all}) for the first shell only, it is found a significant change in the SRO parameter for Mo-Ta pair in the quinary HEAs. The second NN contribution to SRO parameter for the Mo-Ta pair becomes positive and the resulting average SRO parameter, $\alpha _{\text{1+2}}^{\text{MoTa}}$, is now reduced in low temperature region in a comparison with the 1NN SRO parameter, $\alpha_{\text{2,1}}^{\text{MoTa}}$. This confirms that in equimolar Mo-Nb-Ta-V-W HEAs, the presence of $B2$ phase where the 2NN shell contain only atoms of the same chemical species for Mo-Mo or Ta-Ta pairs are very favorable from the short-range order consideration. A similar reduction in average SRO parameter for the Mo-Ta pair is also consistently predicted in the quaternary alloys without the presence of Nb element (see Fig.(\ref{fig:SRO_NNN_without_Nb})).

An other interesting result found by comparing Figs. (\ref{fig:SRO_NNN}) and (\ref{fig:SRO_all}) is that when the 2NN contribution is taken into consideration, the average SRO parameters for the V-W pair become more dominant and comparable with the corresponding values for Mo-W. By analyzing separately the contributions from the 1NN and 2NN, it is found that the 2NN SRO parameter $\alpha_{\text{2,2}}^{\text{VW}}$ for V-W pair is negative in a large rage of temperatures and only becomes positive at very low temperatures smaller than 100K. Therefore unlike Mo-Ta binary, the formation of $B2$-like phase for V-W bcc-binary systems in the considered quinary HEA system is therefore only favorable at the low-temperature phase transition. For temperatures higher than 100K the negative value of SRO parameter indicates the presence of unlike V-W pairs in the 2NN. The latter finding seems to support the local chemical environment of the ordered $B32$ phase which is the ground-state structure predicted by DFT calculations in the equimolar binary system [\onlinecite{Muzyk2011}]. A similar analysis for average SRO parameter for V-W pair in the two quaternary sub-systems without Nb and Ta (see Fig.\ref{fig:SRO_NNN_without_Nb} and \ref{fig:SRO_NNN_without_Ta}, respectively) shows that the phase transition from the ordered-like $B32$ configuration to $B2$-like structure occurs at the temperatures smaller than 200K.   

Finally, by including the 2NN effects, the average SRO $\alpha _{\text{1+2}}^{\text{TaV}}$ for Ta-V pair (see Fig.\ref{fig:SRO_NNN_without_Nb}) and $\alpha _{\text{1+2}}^{\text{NbV}}$ - for Nb-V pair (see Fig.\ref{fig:SRO_NNN_without_Ta}) are now clearly positive at the very low-temperature region of quaternay HEAs Mo-Ta-V-W and Mo-Nb-V-W, respectively. Both of these SRO parameters remain strongly positive in quinary Mo-Nb-Ta-V-W system (see (\ref{fig:SRO_NNN})) showing the segregation trend between Ta and V as well as between Nb and V. This predicted trend is consistent with atomic configuration visualized in Fig.\ref{fig:atomistic_picture_400K}.

\section{Conclusion}

A theoretical formulation for the Warren-Cowley short-range order parameters has been systematically developed for the five and four-component alloy systems. Their analytical expressions are written in terms of correlation functions which can be computed from many-body cluster-expansion Hamiltonian in a combination with Monte Carlo simulations using ECIs. As an application of the formulation, the dependence of SRO as a function of temperature has been analyzed consistently for the equimolar quinary Mo-Nb-Ta-V-W and two of its quaternary sub-systems (Mo-Nb-V-W and Mo-Ta-V-W) in bcc lattice. It is found that below temperatures of 750 K there is an order-disorder phase transition from the high-temperature limit of ideal solutions to a non-random distribution of different chemical species where the excess of configuration entropy terms come from non-zero values of SRO parameters. In general, the SRO exists in HEAs structures that show a preference of a particular pair of atoms to occur as first neighbors. In the present study, a strong negative SRO parameter has been predicted for Mo-Ta pair in quinary Mo-Nb-Ta-V-W and quaternary Mo-Ta-V-W HEAs. The origin of this phenomenon comes from the negative enthalpy of mixing and therefore the effective first-nearest neighbor interaction between Mo and Ta in the bcc binary system in our DFT data base calculations for the CE Hamiltonian. It is also in an agreement with previous ab-initio analysis for phase transition between the disordered ($A2$) to the ordered ($B2$) phases [\onlinecite{Widom2013}] as well as electronic structure calculations for the quaternary Mo-Nb-Ta-W system [\onlinecite{Sluiter2017}]. For the case of quaternary Mo-Nb-V-W HEAs where there are no Ta atoms, it is found that the chemical preference for a Mo-Nb pair becomes dominant in the first nearest neighbor and therefore its SRO is strongly negative at low temperature. For all the three considered HEAs, the chemical SRO parameter for W and V pair is also predicted to be strongly negative and consistently preferable in the first neighbor of the bcc lattice. It is important to note that the values of three SRO parameters, $\alpha_{\text{2,1}}^{\text{MoTa}}$, $\alpha_{\text{2,1}}^{\text{MNb}}$, $\alpha_{\text{2,1}}^{\text{WV}}$, represent the chemical bonding of transition metals between the group V and VI in the periodic table of elements and their negative values can be understood from the analysis of mixing enthalpy in the corresponding binary systems. 
        
The analytically calculated Warren-Cowley SRO parameters for other pairs in the equimolar quinary and quaternary systems are positive at low temperature showing different degrees of phase segregation between them in first nearest neighbors. While the positive values of $\alpha_{\text{2,1}}^{\text{MoW}}$, $\alpha_{\text{2,1}}^{\text{NbTa}}$, $\alpha_{\text{2,1}}^{\text{NbV}}$ and $\alpha_{\text{2,1}}^{\text{TaV}}$ can be explained from the fact that thermodynamically the chemical bonding is not favorable for bcc binary between transition metal elements from the same group V or VI, the separation between Nb-W, Ta-W and Mo-V pairs in HEAs are not directly related to the mixing enthalpy values of their binaries. The complex trend of SRO parameters in multi-component alloys therefore can only be explained from the cluster-expansion Hamiltonian model which goes beyond the nearest-neighbor pair-wise approximation in the present study.

A further investigation of SRO parameters by including the 2NN shell in bcc alloys confirms the formation of favorable $B2$ phase for the Mo-Ta binaries in both quinary and quaternary systems. The investigation for V-W pair shows that the average SRO parameter becomes comparably negative to those for Mo-Ta pair.The formation of $B2$-like, where the 1NN and 2NN SRO parameters have the opposite signs,  is only favorable for the V-W in the considered HEAs at very the low temperatures, namely lower than 100K for the quinary and 200K for the quaternary systems. At higher temperatures, both 1NN and 2NN SRO parameters for V-W pair are found to be negative indicating a similar local environment of the ground-state $B32$ phase which has been predicted previously in the corresponding bcc binary system. The average SRO parameters for both 1NN and 2NN pairs confirm a clear trend of segregation between Nb and V (or Ta and V) elements from the group V of bcc transition metal series. 

Finally, it is noted that the SRO phenomenon considered in this work describes the degree of local deviation from the average on a local scale in term of chemical occupation. The displacive deviation due to atomic size effects is not considered here although a generalized theory of SRO from first-principles calculations can be taken into account [\onlinecite{Toda-Caraballo2015,Wolverton1998,Widom2015}]. It would be very desirable to compare the present theoretical data for SRO in Mo-Nb-Ta-V-W with a detailed experimentally detectable analysis of local chemical order, for example, from neutron diffraction experiment first produced in [\onlinecite{Senkov2011}].

\section{Acknowledgements}

This work has been carried out within the framework of the EUROfusion Consortium and has received funding from the Euratom research and training programme 2014-2018 under grant agreement No 633053 and funding from the RCUK Energy Programme [grant number EP/P012450/1].  The views and opinions expressed herein do not necessarily reflect those of the European Commission. AFC acknowledges financial support from EPSRC (EP/L01680X/1) through the Materials for Demanding Environments Centre for Doctoral Training.
JSW acknowledges the financial support from the Foundation of Polish Science grant HOMING (No. Homing/2016-1/12). The HOMING programme is co-financed by the European Union under the European Regional Development Fund. The simulations were partially carried out by JSW with the support of the Interdisciplinary Centre for Mathematical and Computational Modelling (ICM), University of Warsaw, under grant no GA65-14. DNM would like to acknowledge the support from Marconi-Fusion, the High Performance Computer at the CINECA headquarters in Bologna (Italy) for its provision of supercomputer resources.

\bibliography{SRO_HEA}
\end{document}